\documentclass{ws-procs9x6}

\def\to{\rightarrow}

\def\PL{{Phys.\ Lett.\ }}

\def\PRD{{Phys.\ Rev.\ D} }
\def\PRL{{Phys.\ Rev.\ Lett.\ }}

\def\al{\alpha}

\def\de{\delta}

\def\ve{\varepsilon}

\def\th{\theta}

\def\ka{\kappa}

\def\rh{\rho}

\def\si{\sigma}

\def\cl{{\mathcal L}}

\def\fr#1#2{{{#1} \over {#2}}}
\def\half{{\textstyle{1\over 2}}}

\def\frac#1#2{{\textstyle{{#1}\over {#2}}}}

\def\lsim{\mathrel{\rlap{\lower4pt\hbox{\hskip1pt$\sim$}}
    \raise1pt\hbox{$<$}}}
\def\gsim{\mathrel{\rlap{\lower4pt\hbox{\hskip1pt$\sim$}}
    \raise1pt\hbox{$>$}}}
\def\sqr#1#2{{\vcenter{\vbox{\hrule height.#2pt
         \hbox{\vrule width.#2pt height#1pt \kern#1pt
         \vrule width.#2pt}
         \hrule height.#2pt}}}}

\def\prt{\partial}
\def\lrpartial{\raise 1pt\hbox{$\stackrel\leftrightarrow\partial$}}

\def\etal{{\it et al.}}

\newcommand{\beq}{\begin{equation}}
\newcommand{\eeq}{\end{equation}}
\newcommand{\bea}{\begin{eqnarray}}
\newcommand{\eea}{\end{eqnarray}}
\newcommand{\rf}[1]{(\ref{#1})}

\begin{document}

\title{Spacetime symmetries and varying scalars}

\author{RALF LEHNERT}

\address{Department of Physics and Astronomy\\
Vanderbilt University,
Nashville, Tennessee, 37235\\ 
E-mail: ralf.lehnert@vanderbilt.edu}

\maketitle

\abstracts{This talk discusses the relation 
between spacetime-dependent scalars,
such as couplings or fields,
and the violation of Lorentz symmetry.
A specific cosmological supergravity model
demonstrates 
how scalar fields can acquire time-dependent expectation values.
Within this cosmological background,
excitations of these scalars
are governed 
by a Lorentz-breaking dispersion relation.
The model also contains couplings of the scalars
to the electrodynamics sector
leading to the time dependence of both
the fine-structure parameter $\alpha$
and the $\theta$ angle.
Through these couplings,
the variation of the scalars
is also associated
with Lorentz- and CPT-violating effects
in electromagnetism.
}

\section{Introduction}

Despite its phenomenological success,
the Standard Model of particle physics
leaves unresolved a variety of theoretical issues.
Substantial experimental and theoretical efforts
are therefore directed 
toward the search 
for a more fundamental theory 
that includes a quantum description of gravity.
However, 
most quantum-gravity effects 
in virtually all leading candidate models 
are expected to be minuscule
due to Planck-scale suppression.

Recently,
minute violations of Lorentz and CPT symmetry
have been identified as promising
Planck-scale signals.\cite{cpt01}
The idea is 
that these symmetries hold exactly in established physics,
are accessible to ultrahigh-precision tests, 
and can be broken in various quantum-gravity candidates.
As examples, 
we mention
strings,\cite{kps}
spacetime foam,\cite{ell98,suv}
nontrivial spacetime topology,\cite{klink}
loop quantum gravity,\cite{amu}
and noncommutative geometry.\cite{chklo}

The emerging low-energy effects of Lorentz and CPT breaking
are described by the Standard-Model Extension (SME).\cite{sme}
The SME is a field-theory framework 
at the level of the usual Standard Model
and general relativity.
Its flat-spacetime limit 
has provided the basis 
for numerous experimental and theoretical studies
of Lorentz and CPT violation
involving 
mesons,\cite{hadronexpt,kpo,hadronth,ak}
baryons,\cite{ccexpt,spaceexpt,cane}
electrons,\cite{eexpt,eexpt2,eexpt3}
photons,\cite{photon}
muons,\cite{muons}
and the Higgs sector.\cite{higgs}
We remark 
that neutrino-oscillation experiments
offer the potential for discovery.\cite{sme,neutrinos,nulong}

Varying scalars 
are another feature of many approaches to
fundamental physics.
Effective couplings, 
for instance, 
typically acquire time dependencies 
in models with extra dimensions.\cite{theo}
Another class of models 
contains scalar fields, 
which can acquire time-dependent expectation values
driven by the expansion of the universe.
For example,
in modern approaches to cosmology,
such as quintessence,\cite{quint} 
k essence,\cite{kessence} 
or inflation,\cite{infl}
scalar fields are frequently invoked
to explain certain observations.

In the present talk,
it is demonstrated
that the above potential quantum-gravity features
are interconnected.
In particular,
spacetime-dependent scalars
are typically associated 
with Lorentz and possibly CPT violation.
In Sec.\ \ref{arg},
general arguments in favor 
of this claim
are given.
For further illustrations
and some specific results,
a toy model 
is introduced
in Sec.\ \ref{models}.
Lorentz violating effects
within the scalar sector
of our toy cosmology
are discussed in Sec.\ \ref{scalar}.
Section \ref{scalcoupl} 
discusses the Lorentz and CPT breaking
in the electrodynamics sector
of the model. 
Section \ref{sum}
contains a brief summary.

\section{General arguments}
\label{arg}

A spacetime-dependent scalar,
regardless of the mechanism driving the variation,
typically implies the breaking of spacetime-translation invariance.
Since translations and Lorentz transformations
are closely linked in the Poincar\'e group,
it is reasonable to expect
that the translation-symmetry violation
also affects Lorentz invariance.

Consider,
for instance,
the angular-momentum tensor $J^{\mu\nu}$,
which is the generator for Lorentz transformations:
\beq
J^{\mu\nu}=\int d^3x \;\big(\th^{0\mu}x^{\nu}-\th^{0\nu}x^{\mu}\big).
\label{gen}
\eeq
Note
that this definition 
contains the energy--momentum tensor $\th^{\mu\nu}$,
which is not conserved
when translation invariance is broken.
In general,
$J^{\mu\nu}$
will possess a nontrivial dependence on time,
so that the usual time-independent 
Lorentz-transformation generators do not exist.
As a result,
Lorentz and CPT symmetry
are no longer assured.

Intuitively,
the violation of Lorentz invariance 
in the presence of a varying scalar can be understood as follows.
The 4-gradient of the scalar must be nonzero
in some regions of spacetime. 
Such a gradient 
then selects a preferred direction in this region.
Consider, 
for example, 
a particle
that interacts with the scalar.
Its propagation features
might be different
in the directions parallel and perpendicular to the gradient.
Physically inequivalent directions
imply the violation of rotation symmetry.
Since rotations are contained in the Lorentz group,
Lorentz invariance must be violated.

Lorentz violation induced by varying scalars
can also be established at the Lagrangian level.
Consider, 
for instance,
a system with varying coupling $\xi(x)$
and scalar fields $\phi$ and $\Phi$,
such that the Lagrangian $\mathcal{L}$ contains a term
$\xi(x)\,\partial^{\mu}\phi\,\partial_{\mu}\Phi$.
The action for this system can be integrated by parts
(e.g., with respect to the first partial derivative in the above term)
without affecting the equations of motion.
An equivalent Lagrangian $\mathcal{L}'$ would then obey
\begin{equation}
\mathcal{L}'\supset -K^{\mu}\phi\,\partial_{\mu}\Phi,
\label{example}
\end{equation}
where $K^{\mu}\equiv\partial^{\mu}\xi$ is an external
nondynamical 4-vector,
which clearly violates Lorentz symmetry.
We remark
that for variations of $\xi$ on cosmological scales,
$K^{\mu}$ is constant to an excellent approximation 
locally---say on solar-system scales. 

\section{Specific cosmological model}
\label{models}

In the remainder of this talk, 
we illustrate the result from the previous section
within a specific supergravity model. 
This model
generates the variation of two scalars $A$ and $B$
in a cosmological context.
It leads to a varying fine-structure parameter $\alpha$
and a varying electromagnetic $\theta$ angle.
The starting point is pure $N=4$ supergravity
in four spacetime dimensions.
Although unrealistic in its details,
it can give qualitative insights into candidate fundamental physics
because it is a limit of  $N=1$ supergravity 
in eleven dimensions,
which is contained in M-theory.

When only one graviphoton $F^{\mu\nu}$
is excited,
the bosonic part of pure $N=4$ supergravity
reads\cite{cj,klp03}
\bea
\ka \cl_{\rm sg}
&=&
-\frac 1 2 \sqrt{g} R
+\sqrt{g} ({\prt_\mu A\prt^\mu A + \prt_\mu B\prt^\mu B})/{4B^2}
\nonumber\\
&&
\qquad\!
-\frac 1 4 \ka \sqrt{g} M F_{\mu\nu} F^{\mu\nu}
-\frac 1 4 \ka \sqrt{g} N F_{\mu\nu} \tilde{F}^{\mu\nu}\; .
\label{lag2}
\eea
Here,
\beq
M =\fr
{B (A^2 + B^2 + 1)}
{(1+A^2 + B^2)^2 - 4 A^2}\; ,
\quad
N = \fr
{A (A^2 + B^2 - 1)}
{(1+A^2 + B^2)^2 - 4 A^2}\; ,
\label{N}
\eeq
$\tilde{F}^{\mu\nu}=\ve^{\mu\nu\rh\si}F_{\rh\si}/2$
denotes the dual field-strength tensor,
and $g=-\det (g_{\mu\nu})$.
We remark
that  the redefinition
$F^{\mu\nu}\to F^{\mu\nu}/\sqrt{\ka}$
removes the explicit appearance 
of the gravitational coupling $\ka$
in the equations of motion.

As a further ingredient,
we gauge the internal SO(4) symmetry
of the full $N=4$ supergravity Lagrangian,
which supports the interpretation of $F^{\mu\nu}$ 
as the electromagnetic field-strength tensor.
This leads to a potential for the scalars $A$ and $B$
that is unbounded from below.\cite{dfr77}
At this point,
we take a phenomenological approach
and assume 
that in a realistic situation
stability must be ensured
by additional fields and interactions.
At first order, 
we can then model
the potential for the scalars
with the following mass-type terms
\beq
\delta\cl=-\half \sqrt{g} (m_A^2 A^2+ m_B^2 B^2)\; ,
\label{lagr}
\eeq
which we add to $\cl_{\rm sg}$ 
in Eq.\ \rf{lag2}.

The full $N=4$ supergravity Lagrangian
also contains fermionic matter.\cite{cj}
In the present cosmological context,
we can effectively represent the fermions
by the energy--momentum tensor $T_{\mu\nu}$ of dust
describing galaxies and other matter:
\beq
T_{\mu\nu} = \rh u_\mu u_\nu \; .
\label{dust}
\eeq
As usual,
$\rh$ is the energy density of the matter
and $u^\mu$ is a unit timelike vector
orthogonal to the spatial hypersurfaces.

We are now ready to search for cosmological solutions
of our supergravity model.
We proceed under the usual assumption 
of an isotropic homogeneous
flat $(k=0)$ Friedmann--Robertson--Walker universe
with the conventional line element
\beq
ds^2 = dt^2 - a^2(t)\; (dx^2 + dy^2 + dz^2)\; .
\label{frw}
\eeq
Here, 
$a(t)$ denotes the scale factor
and $t$ the comoving time.
Since isotropy requires $F^{\mu\nu}=0$
on large scales,
our cosmology is governed
by the Einstein equations
and the equations of motion for the scalars $A$ and $B$.
Note 
that the fermionic matter is uncoupled from the scalars
at tree level,
so that we can take $T_{\mu\nu}$
as covariantly conserved separately.
It then follows that
$\rh(t) = c_{\rm n}/a^3(t)$,
where $c_{\rm n}$ is an integration constant.

\begin{figure}
\centerline{\epsfxsize=4.1in\epsfbox{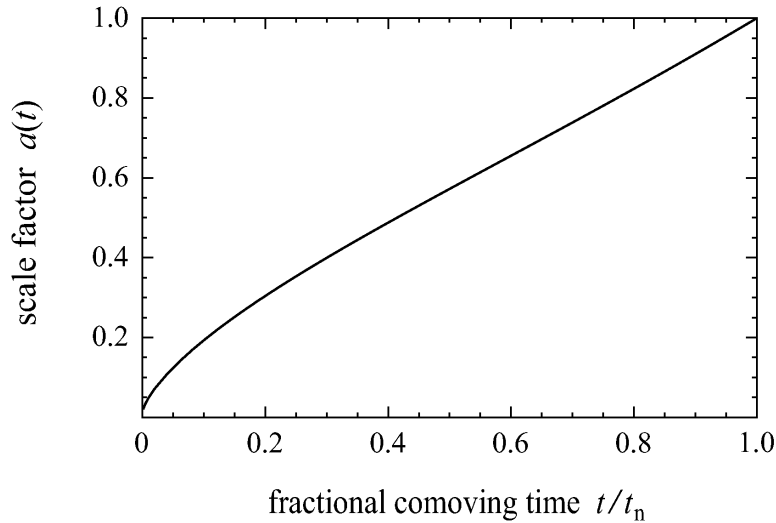}}   
\caption{Scale factor $a(t)$ versus fractional comoving time $t/t_{\rm n}$.
It turns out that the expansion history of this model
matches closely the observed one.
\label{sfsol}}
\end{figure}

\begin{figure}
\centerline{\epsfxsize=4.1in\epsfbox{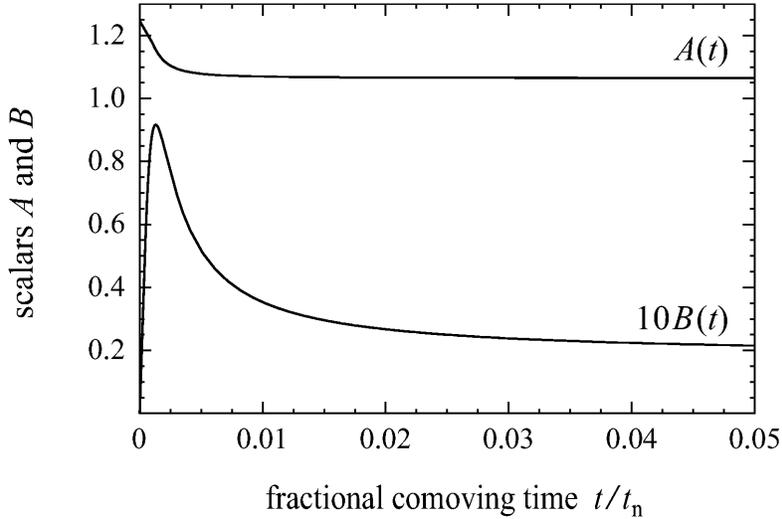}}   
\caption{Time dependence of the scalars $A$ and $B$.
At late times, the scalars approach constant values.
\label{ABsol}}
\end{figure}

Although analytical solutions
within this cosmological model
can be found in special cases,\cite{klp03,blpr}
numerical integration is necessary in general.
A particular solution
is depicted in Figs.\ \ref{sfsol} and \ref{ABsol},
where the following priors
have been used:\cite{blpr}
\bea
m_A & = & 2.7688 \times 10^{-42}\, {\rm GeV}\; ,\nonumber\\
m_B & = & 3.9765 \times 10^{-41}\, {\rm GeV}\; ,\nonumber\\
c_{\rm n} & = & 2.2790 \times 10^{-84}\, {\rm GeV}^2\; ,\nonumber\\
a(t_{\rm n}) & = & 1 \; ,\nonumber\\
A(t_{\rm n}) & = & 1.0220426 \; ,\nonumber\\
\dot{A}(t_{\rm n}) & = & -8.06401\times 10^{-46}\, {\rm GeV}\; ,\nonumber\\
B(t_{\rm n}) & = & 0.016598 \; ,\nonumber\\
\dot{B}(t_{\rm n}) & = & -2.89477\times 10^{-45}\, {\rm GeV}\; .
\label{values}
\eea
Here, 
the dot denotes differentiation
with respect to the comoving time,
and the subscript n indicates the present value of the quantity.
For our present purposes,
the details of this solution are less interesting.
Note,
however,
that the scalars $A$ and $B$ 
have acquired a dependence on the comoving time $t$,
so that they vary on cosmological scales.

Such a  spacetime variation of $A$ and $B$
has various implications
for the scalars themselves
and (due to the coupling of $A$ and $B$ to $F^{\mu\nu}$)
for electrodynamics.
These implications are discussed
in Secs.\ \ref{scalar} and \ref{scalcoupl},
respectively.

\section{Effects in the scalar sector}
\label{scalar}

To gain insight into 
how the time-dependent cosmological background solutions $A_{\rm b}$ and $B_{\rm b}$
affect the scalars themselves,
we investigate the propagation properties
of small localized excitations $\de A$ and $\de B$
of the cosmological background $A_{\rm b}$ and $B_{\rm b}$.
For such a study,
it is appropriate to work in local coordinates,
which we take to be anchored at the spacetime point $x_0$.

Substituting the ansatz
\bea
A(x)&=& A_{\rm b}(x)+\de A(x)\nonumber\\
B(x)&=& B_{\rm b}(x)+\de B(x)
\label{pert}
\eea
for the scalars
into the equations of motion for $A$ and $B$
determines the dynamics
of  the perturbations $\de A$ and $\de B$.
One obtains the following linearized equations:
\bea
{}\!\!\!0 &=& \big[\Box -2B^{\mu}\prt_{\mu}
+2m_A^2B_{\rm b}^2\big]\de A
-\big[2A^{\mu}\prt_{\mu}
-2A^{\mu}B_{\mu}-4m_A^2A_{\rm b}B_{\rm b}\big]\de B\; ,\nonumber\\
0 &=& \big[2A^{\mu}\prt_{\mu}\big]\de A
+\big[\Box  -2B^{\mu}\prt_{\mu}
+6m_B^2B_{\rm b}^2-A^{\mu}A_{\mu}+B^{\mu}B_{\mu}\big]\de B\; .
\label{ablineq}
\eea
Here,
$A_{\rm b}$ and $B_{\rm b}$
as well as 
$A^{\mu}\equiv B_{\rm b}^{-1}\prt^{\mu}A_{\rm b}$,
and $B^{\mu}\equiv B_{\rm b}^{-1}\prt^{\mu}B_{\rm b}$
are evaluated at $x=x_0$.

Equation \rf{ablineq} 
determines the propagation features
of $\de A$ and $\de B$
in the varying cosmological background.
In the context of this equation,
$A^{\mu}$ and $B^{\mu}$
are external nondynamical vectors
selecting a preferred direction
in the local inertial frame.
It follows
that the propagation of $\de A$ and $\de B$
in the varying cosmological background
fails to obey Lorentz symmetry.
This result carries over
to quantum theory:
the traveling disturbances $\de A$ and $\de B$
would be seen as the effective particles
corresponding to the scalars $A$ and $B$,
so that such particles would violate Lorentz invariance.

We remark that the usual ansatz with $\exp(-ip\cdot x)$
yields the plane-wave dispersion relation.
As expected,
this dispersion relation contains
the fixed vectors $A^{\mu}$ and $B^{\mu}$
contracted with the momentum $p^{\mu}$
implying Lorentz breaking.
In addition, 
this dispersion relation 
exhibits imaginary terms 
leading to decaying solutions.
This is consistent
with the nonconservation of 4-momentum
due to the violation of translational invariance.
We also point out
that the equations of motion are coupled,
so that a plane-wave
is a linear combination
of  $\de A$ and $\de B$.
More details can be found in Ref. \refcite{blpr}.

\section{Effects in the scalar-coupled sector}
\label{scalcoupl}

Instead of excitations $\de A$ and $\de B$ of the scalar fields,
we now consider
excitations of $F_{\mu\nu}$
in our background cosmological solution
$A_{\rm b}$ and $B_{\rm b}$.
Again, 
it is appropriate to work in a local inertial frame.
Then, 
the effective Lagrangian $\cl_{\rm cosm}$ for localized $F_{\mu\nu}$ fields
follows from Eq.\ \rf{lag2}
\beq
\cl_{\rm cosm}=
-\frac 1 4 M_{\rm b}F_{\mu\nu} F^{\mu\nu}
-\frac 1 4 N_{\rm b} F_{\mu\nu} \tilde{F}^{\mu\nu}\; ,
\label{efflag}
\eeq
where $M_{\rm b}$ and $N_{\rm b}$
are determined 
by the time-dependent cosmological solutions $A_{\rm b}$ and $B_{\rm b}$ 
for the scalars. 

\begin{figure}
\centerline{\epsfxsize=4.1in\epsfbox{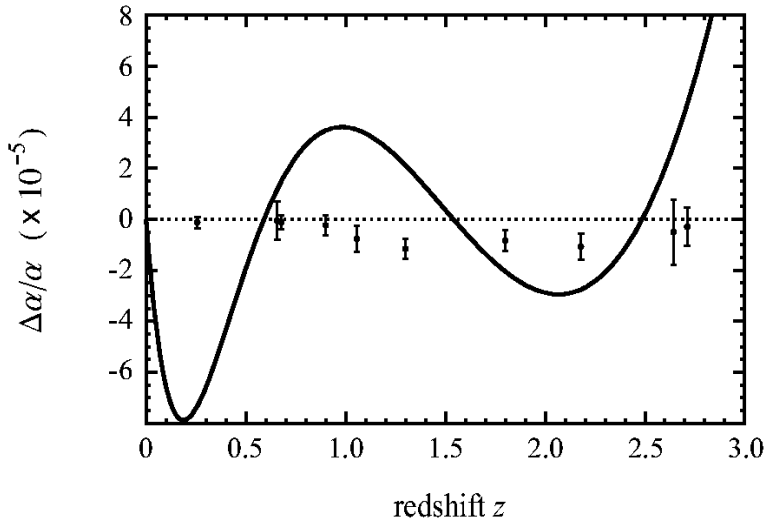}}   
\caption{Relative change in the fine-structure parameter $\alpha$
versus redshift. 
For comparison, 
the Webb data has been included
into the plot.
Although disagreeing in detail,
the data and the model 
exhibit variations of $\alpha$ 
of roughly the same order of magnitude.
\label{alphaev}}
\end{figure}

The physics content of $\cl_{\rm cosm}$
is best extracted 
by comparison
to the conventional electrodynamics Lagrangian $\cl_{\rm em}$
given by
\beq
\cl_{\rm em} =
-\fr{1}{4 e^2} F_{\mu\nu}F^{\mu\nu}
- \fr{\th}{16\pi^2} F_{\mu\nu} \tilde{F}^{\mu\nu}\; .
\label{em}
\eeq
This  shows
that $e^2 \equiv 1/M_{\rm b}$ and  $\th \equiv 4\pi^2 N_{\rm b}$.
Since $M_{\rm b}$ and $N_{\rm b}$
are functions of the varying scalar background $A_{\rm b}$ and $B_{\rm b}$,
the electromagnetic couplings $e$ and $\th$
are no longer constant in general.
In other words,
in our supergravity cosmology
the fine-structure parameter $\al$ and the electromagnetic $\th$ angle
acquire related spacetime dependences.
The evolution of $\al$ 
in the present model with initial conditions \rf{values}
is depicted in Fig.\ \ref{alphaev}. 
Also shown is the recently reported Webb dataset\cite{Webb}
obtained from measurements of high-redshift spectra.

Lorentz violation in our effective electrodynamics 
can be clearly established
by inspection of the modified Maxwell equations
resulting from the Lagrangian \rf{efflag}:
\beq
\fr{1}{e^2}\partial^{\mu}F_{\mu\nu}
-\fr{2}{e^3}(\partial^{\mu}e)F_{\mu\nu}
+\fr{1}{4\pi^2}(\partial^{\mu}\th)\tilde{F}_{\mu\nu}=0\; .
\label{Feom}
\eeq
In our supergravity cosmology,
the gradients of $e$ and $\th$
appearing in Eq.\ \rf{Feom}
are nonzero,
approximately constant in local inertial frames,
and act as a nondynamical external background.
This vectorial background
selects a preferred direction
in the local inertial frame
violating Lorentz symmetry.

We remark
that the term exhibiting the gradient of $\th$
can be identified 
with a Chern--Simons-type contribution to the modified Maxwell equations.
Such a term,
which is contained in the minimal SME,
has received a lot of attention recently.\cite{mcs}
For example,
it typically leads to vacuum \v{C}erenkov radiation.\cite{cer}

\section{Summary}
\label{sum}

In this talk, 
it has been demonstrated
that the violation of spacetime-translation invariance
is closely intertwined 
with the breaking of Lorentz symmetry.
More specifically,
a varying scalar---regardless of the mechanism driving the variation---is associated
with a nonzero gradient,
which selects a preferred direction in spacetime.
This mechanism for Lorentz violation
is interesting because
varying scalars appear in many cosmological contexts.

\section*{Acknowledgments}
Funding by the Funda\c{c}\~ao para a Ci\^encia e a Tecnologia (Portugal)
under Grant No.\ POCTI/FNU/49529/2002 
and by the Centro Multidisciplinar de Astrof\'{\i}sica (CENTRA)
is gratefully acknowledged.

\end{document}